\documentclass[aps,prc,preprint,superscriptaddress,showpacs,floatfix,nobibnotes]{revtex4}
\usepackage{mathrsfs}

\usepackage{bm}
\usepackage{graphicx}
\usepackage{longtable}

\usepackage{graphics}
\usepackage{amsmath}

\begin{document}

\title{Mass of \emph{Y}(4140) in Bethe-Salpeter equation for quarks }

\author{Xiaozhao Chen}\email{chen_xzhao@sina.com} \email[corresponding author]{}
\affiliation{Department of Foundational courses, Shandong University
of Science and Technology, Taian, 271019, China}

\author{Xiaofu L\"{u}}
\affiliation{Department of Physics, Sichuan University, Chengdu,
610064, China}

\author{Renbin Shi}
\affiliation{Department of Foundational courses, Shandong University
of Science and Technology, Taian, 271019, China}

\author{Xiurong Guo}
\affiliation{Department of Foundational courses, Shandong University
of Science and Technology, Taian, 271019, China}

\date{\today}

\begin{abstract}
Using the general form of the Bethe-Salpeter wave functions for the bound states consisting of
two vector fields given in our previous work, we investigate the molecular state composed of $D^{*+}_s$$D^{*-}_s$. However, for the SU(3) symmetry
the component $D^{*+}_s$$D^{*-}_s$ is coupled with the other components $D^{*0}$$\bar{D}^{*0}$ and $D^{*+}$$D^{*-}$. Then we interpret the internal structure of the observed \emph{Y}(4140) state as a mixed state of pure molecule states $D^{*0}$$\bar{D}^{*0}$, $D^{*+}$$D^{*-}$ and $D^{*+}_s$$D^{*-}_s$ with quantum numbers $J^P=0^+$. In this paper, the operator product expansion is used to introduce the nonperturbative contribution from the vacuum condensates into the interaction between two heavy mesons. The calculated mass of \emph{Y}(4140) is consistent with the experimental value, and we conclude that it is a more reasonable scenario to explain the structure of Y (4140) as a mixture of pure molecule states.
\end{abstract}



\maketitle

\newpage

\parindent=20pt

\section{Introduction}
\label{intro}
The narrow state \emph{Y}(4140) was discovered by CDF
collaboration \cite{T.Aaltonen} and its structure does not fit the conventional $c\bar c$
charmonium interpretation. Then possible interpretations beyond quark-antiquark state have
been proposed, such as hadronic molecule state \cite{liu} and tetraquark state \cite{wang}. Following the CDF result,
it is suggested in Ref. \cite{liu} that the \emph{Y}(4140) is a molecular state of
$D^{*+}_sD^{*-}_s$. However, there are some defects in Ref. \cite{liu}: the
numerical result sensitively depends on the adjustable parameter, the heavy vector mesons
$D^{*+}_s$ and $D^{*-}_s$ are considered as pointlike objects, and the definite spin-parity quantum numbers $J^P$
of the \emph{Y}(4140) can not be deduced in theory. More importantly,
the previous work \cite{liu} dealt with this two-body system $D^{*+}_sD^{*-}_s$ in the formalism
of quantum mechanics and the potential between two heavy mesons was constructed in perturbation theory.
Therefore the nonperturbative effects in quantum chromodynamics (QCD), for example,
the condensates of vacuum can not be considered in their work.

In quantum field theory, the most general form of the Bethe-Salpeter (BS) wave functions
for the bound states composed of two vector fields of arbitrary
spin and definite parity has been given \cite{mypaper4}. In this work, we apply the general
formalism to investigate the molecular state of $D^{*+}_sD^{*-}_s$ and consider that the effective interaction between these two heavy mesons is derived
from one light vector meson exchange. In Ref. \cite{mypaper4}, we have deduced that in our approach one light pseudoscalar meson exchange has no contribution to the potential between two heavy vector mesons. Because of the SU(3) symmetry of the light vector mesons, one strange meson ($K^*$) exchange should be considered. From one strange meson exchange,
it is necessary to consider the mixing of three  pure molecule states $D^{*0}$$\bar{D}^{*0}$, $D^{*+}$$D^{*-}$ and $D^{*+}_sD^{*-}_s$, which is not considered in Ref. \cite{liu}. Therefore, we assume that the \emph{Y}(4140) state is a combination of three pure molecule states $D^{*0}$$\bar{D}^{*}$, $D^{*+}$$D^{*-}$ and $D^{*+}_s$$D^{*-}_s$.

To construct the interaction kernels between two heavy vector mesons $D^{*0}$$\bar{D}^{*}$, $D^{*+}$$D^{*-}$ and $D^{*+}_s$$D^{*-}_s$, we still consider that the heavy vector meson is a
bound state composed of a light quark and c-quark and investigate the interaction of the light meson with the light quark in the heavy meson,
which will be emphatically reconsidered in this work. As well known, the nonperturbative contribution plays an important role in the case of QCD at low energy. It is necessary to note that the more nonperturbative effects should be taken into account when we investigate the light meson interaction with quark in the heavy meson. In this work, we introduce the operator product expansion and obtain the heavy meson BS wave function including the contribution from the condensates of vacuum. From the improved heavy meson BS wave function, we can obtain the heavy meson form factors which contain the contribution from the nonperturbative effects of QCD. Through these further form factors we can obtain the heavy meson interaction with light meson and the potentials between two heavy vector mesons without an extra parameter. Obviously, this approach is closer to QCD than our previous works \cite{mypaper4,mypaper,mypaper2,mypaper3}. Then numerically solving the relativistic Schr$\ddot{o}$dinger-like equation with these potentials, we obtain the wave functions of the pure molecule states $D^{*0}$$\bar{D}^{*}$, $D^{*+}$$D^{*-}$ and $D^{*+}_s$$D^{*-}_s$, respectively. Finally, using the coupled-channel approach, we can obtain the masses and the wave functions of the mixed states and then the definite
quantum numbers of the \emph{Y}(4140) system can be deduced.

This paper is organized in the following way. In Sec. \ref{sec:RGFBSWF} we show the general form
of BS wave functions for the bound states consisting of two vector fields.
After constructing the interaction kernel between two heavy vector mesons, we introduce the mixed state of three pure molecule states. Sec. \ref{sec:instanapp} shows the procedure of the instantaneous
approximation. Sec. \ref{sec:formfac} shows how to obtain the heavy meson BS wave function and form factor
which contain the contribution from the vacuum condensates. Then the interaction potentials between two heavy vector mesons and the
masses of the pure molecule states are calculated. Sec. \ref{sec:nr} gives the numerical result and our conclusion is presented in
Section \ref{sec:concl}.

\section{The Mixing Mechanism}\label{sec:RGFBSWF}
\begin{figure*}
\centering
\resizebox{0.35\textwidth}{!}{%
  \includegraphics{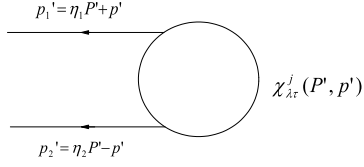}
}
\caption{\label{Fig1} Bethe-Salpeter wave function for the bound
state composed of two vector fields.}
\end{figure*}

If a bound state with spin $j$ and parity $\eta_{P'}$ is composed of two vector fields with masses $M_1$ and $M_2$, respectively,
the BS wave function of this bound state in the momentum representation should satisfy, \cite{mypaper4} for $\eta_{P'}=(-1)^j$,
\begin{equation}\label{jp0}
\chi_{\lambda\tau}^{j=0}(P',p')=T^1_{\lambda\tau}\phi_1+T^2_{\lambda\tau}\phi_2,
\end{equation}
\begin{equation}\label{jp}
\chi_{\lambda\tau}^{j\neq0}(P',p')=\eta_{\mu_1\cdots\mu_j}[p'_{\mu_1}\cdots
p'_{\mu_j}(T^1_{\lambda\tau}\phi_1+T^2_{\lambda\tau}\phi_2)+T^3_{\lambda\tau}\phi_3+T^4_{\lambda\tau}\phi_4],
\end{equation}
and, for $\eta_{P'}=(-1)^{j+1}$,
\begin{equation}\label{jm0}
\chi_{\lambda\tau}^{j=0}(P',p')=\epsilon_{\lambda\tau\xi\zeta}p'_\xi
P'_\zeta\psi_1,
\end{equation}
\begin{equation}\label{jm}
\chi_{\lambda\tau}^{j\neq0}(P',p')=\eta_{\mu_1\cdots\mu_j}(p'_{\mu_1}\cdots
p'_{\mu_j}\epsilon_{\lambda\tau\xi\zeta}p'_\xi P'_\zeta\psi_1+T^5_{\lambda\tau}\psi_2+T^6_{\lambda\tau}\psi_3+T^7_{\lambda\tau}\psi_4+T^8_{\lambda\tau}\psi_5),
\end{equation}
where $P'$ is the bound state momentum, $p'$ is the relative momentum of two vector fields, $\eta_{\mu_1\cdots\mu_j}$ is the polarization tensor describing the spin of the bound state, the independent tensor structures $T^i_{\lambda\tau}$ are given in Appendix A, $\phi_i(P'\cdot p',p'^2)$ and $\psi_i(P'\cdot p',p'^2)$ are independent scalar functions. In this work, we have $P'=p_1'+p_2', p'=\eta_2p_1'-\eta_1p_2'$, $\eta_{1,2}$ are two positive quantities such that $\eta_{1,2}=M_{1,2}/(M_1+M_2)$, $p_1'$ and $p_2'$ are the momenta of two vector fields, respectively. This is presented in Fig. \ref{Fig1}.

In experiment the quantum numbers $J^P$ of the \emph{Y}(4140) are not unambiguously determined except for $C=+$. Assuming that the \emph{Y}(4140) is a S-wave molecule state
whose constituents are two heavy vector mesons $D^{*+}_s$ and $D^{*-}_s$, one can have $J^P=0^+$ or $2^+$ for this system \cite{liu}. From
Eqs. (\ref{jp0}) and (\ref{jp}), the BS wave function
of this bound state becomes, for $J^P=0^+$,
\begin{equation}\label{SBSWF}
\chi_{\lambda\tau}^{0^+}(P',p')=T^1_{\lambda\tau}\mathcal
{F}_1+T^2_{\lambda\tau}\mathcal
{F}_2,
\end{equation}
or, for $J^P=2^+$,
\begin{equation}\label{TBSWF}
\chi_{\lambda\tau}^{2^+}(P',p')=\eta_{\mu_1\mu_2}[p'_{\mu_1}
p'_{\mu_2}(T^1_{\lambda\tau}\mathcal {G}_1+T^2_{\lambda\tau}\mathcal{G}_2)+T^3_{\lambda\tau}\mathcal{G}_3+T^4_{\lambda\tau}\mathcal{G}_4],
\end{equation}
which satisfies the BS equation
\begin{equation}\label{BSE1}
\chi_{\lambda\tau}(P',p')=\int
\frac{id^4q'}{(2\pi)^4}\Delta_{F\lambda\alpha}(p_1')\mathcal
{V}_{\alpha\theta,\beta\kappa}(p',q';P')\chi_{\theta\kappa}(P',q')\Delta_{F\beta\tau}(p_2'),
\end{equation}
where $\mathcal{V}_{\alpha\theta,\beta\kappa}$ is the interaction
kernel, $\Delta_{F\lambda\alpha}(p_1')$ and $\Delta_{F\beta\tau}(p_2')$ are the propagators for the spin 1 fields,
$\Delta_{F\lambda\alpha}(p_1')=(\delta_{\lambda\alpha}+\frac{p'_{1\lambda}
p'_{1\alpha}}{M_1^2})\frac{1}{p_1'^2+M_1^2-i\epsilon}$,
$\Delta_{F\beta\tau}(p_2')=(\delta_{\beta\tau}+\frac{p'_{2\beta}
p'_{2\tau}}{M_2^2})\frac{1}{p_2'^2+M_2^2-i\epsilon}$ and the momentum of this bound
state is set as $P'=(0,0,0,iM)$ in the rest frame. In this approach, we find that one light pseudoscalar meson exchange has no contribution to the potential between two heavy vector mesons \cite{mypaper4}. Then we consider that the effective interaction between $D^{*+}_s$ and $D^{*-}_s$ is derived from one light vector meson exchange. The charmed meson $D^{*+}_s$ is composed of a heavy quark $c$ and a light antiquark $\bar s$. From the SU(3) symmetry, the Lagrangian for the interaction of light vector meson with quarks should be
\begin{equation}\label{Lag}
\mathscr{L}_I=ig_0\left(\begin{array}{ccc} \bar{u}&\bar{d}&\bar{s}
\end{array}\right)\gamma_\mu\left(\begin{array}{ccc}
\rho^0+\frac{1}{\sqrt{3}}V_8+V_1&\sqrt{2}\rho^+&\sqrt{2}K^{*+}\\\sqrt{2}\rho^-&-\rho^0+\frac{1}{\sqrt{3}}V_8+V_1&\sqrt{2}K^{*0}\\\sqrt{2}K^{*-}&\sqrt{2}\bar{K}^{*0}&-\frac{2}{\sqrt{3}}V_8+V_1
\end{array}\right)\left(\begin{array}{c} u\\d\\s
\end{array}\right),
\end{equation}
where the flavor-SU(3) singlet $V_1$ and octet $V_8$ states of vector mesons mix to form the physical $\omega$ and $\phi$ mesons as
\begin{equation}\label{mix}
\phi=-V_8cos\theta+V_1sin\theta,~~~~~
\omega=V_8sin\theta+V_1cos\theta.
\end{equation}
Because of the SU(3) symmetry, we consider that the exchanged mesons should be the singlet $V_1$ and octet $V_8$ states.

\begin{figure*}
\centering
\resizebox{0.8\textwidth}{!}{%
  \includegraphics{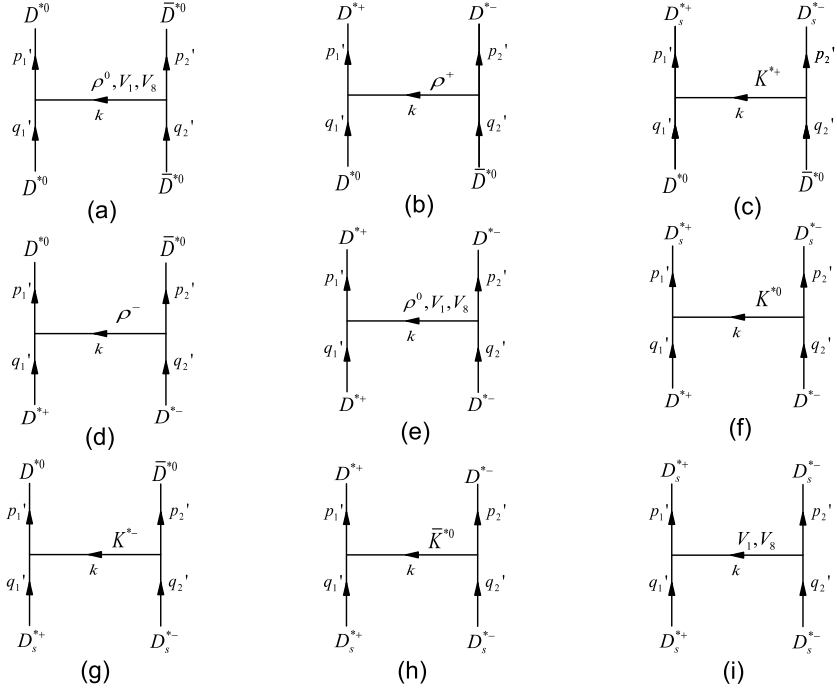}
}
\caption{\label{Fig2} The one light vector meson exchange.}
\end{figure*}

The Lagrangian expressed as Eq. (\ref{Lag}) gives nine S-matrix elements, as shown in Fig. \ref{Fig2}. The graphs (a), (e), (i) in Fig. \ref{Fig2} represent pure molecule states $D^{*0}$$\bar{D}^{*}$, $D^{*+}$$D^{*-}$ and $D^{*+}_s$$D^{*-}_s$, respectively; and the remaining graphs represent the coupled-channel terms between two pure molecule states. Then the observed \emph{Y}(4140) state can not be considered as a pure molecule state of $D^{*+}_s$$D^{*-}_s$ and we interpret it as a mixed state of pure molecule states $D^{*0}$$\bar{D}^{*0}$, $D^{*+}$$D^{*-}$ and $D^{*+}_s$$D^{*-}_s$. The BS wave function of the \emph{Y}(4140) state is a linear combination of these three pure molecule states as
\begin{equation}\label{mixBSwf}
\chi^{Y(4140)}_{\lambda\tau}=\sum_\alpha a_\alpha\chi^{D^{*0}\bar D^{*0}}_{\alpha,\lambda\tau}+\sum_\beta a_\beta\chi^{D^{*+}D^{*-}}_{\beta,\lambda\tau}+\sum_\gamma a_\gamma\chi^{D_s^{*+}D_s^{*-}}_{\gamma,\lambda\tau},
\end{equation}
where $\chi^{D^{*0}\bar D^{*0}}_{\alpha,\lambda\tau}$, $\chi^{D^{*+}D^{*-}}_{\beta,\lambda\tau}$, and $\chi^{D_s^{*+}D_s^{*-}}_{\gamma,\lambda\tau}$ are the eigenstates of Hamiltonian without considering the coupled-channel terms, and these eigenstates have the same quantum numbers. While the pure molecule states $D^{*0}$$\bar{D}^{*0}$ and $D^{*+}$$D^{*-}$ have been investigated in Ref. \cite{mypaper4}, we investigate the pure molecule state $D^{*+}_s$$D^{*-}_s$ as follow.

Now, we construct the interaction kernel between $D^{*+}_s$ and $D^{*-}_s$ from one light vector meson ($V_1$ and $V_8$) exchange, shown as Fig. \ref{Fig2}(i). The contribution of Fig. \ref{Fig2}(i) is only from the terms $ig_1\bar{s}\gamma_\mu V_1s$ and $ig_8\bar{s}\gamma_\mu V_8s$ in Eq. (\ref{Lag}), where $g_1$ is the singlet-quark coupling constant and $g_8$ is the octet-quark coupling constant. From Eq. (\ref{mix}), we obtain the relations of $g_1$ and $g_8$
\begin{equation}
g_\phi=-g_8cos\theta+g_1sin\theta,~~~~~
g_\omega=g_8sin\theta+g_1cos\theta,
\end{equation}
where the meson-quark
coupling constants $g_\omega^2=2.42$ and $g_\phi^2=13.0$ were obtained within QCD sum rules approach
\cite{cc2}, and the mixing angle $\theta=38.58^\circ$ was obtained by KLOE\cite{mix}. Since the SU(3) is broken, the masses of the singlet
$V_1$ and octet $V_8$ states are approximatively identified with the masses of two physically observed $\omega$ and $\phi$ mesons, respectively. Then the effective quark current is $J_\mu=i\bar{s}\gamma_\mu s$ and the S-matrix element between the heavy vector mesons is
\begin{equation}\label{effinter}
\begin{split}
V^I=&g_1^2\langle
V(p_1')|J_\mu|V(q_1')\rangle\bigg(\delta_{\mu\nu}+\frac{k_\mu
k_\nu}{m_\omega^2}\bigg)\frac{1}{k^2+m_\omega^2}\langle
V(p_2')|J_\nu|V(q_2')\rangle\\
&+g_8^2\langle
V(p_1')|J_\mu|V(q_1')\rangle\bigg(\delta_{\mu\nu}+\frac{k_\mu
k_\nu}{m_\phi^2}\bigg)\frac{1}{k^2+m_\phi^2}\langle
V(p_2')|J_\nu|V(q_2')\rangle,
\end{split}
\end{equation}
where $\langle V|J_\mu|V\rangle$ is the vertex of the heavy vector meson interaction with the light vector meson. From the Lorentz-structure, the matrix elements of quark current can be expressed as
\begin{equation}\label{meofqc1}
\begin{split}
\langle V(p_1')|J_{\mu}|V(q_1')
\rangle=&\frac{1}{2\sqrt{E_1(p_1')E_1(q_1')}}\bigg\{[\varepsilon^*(p_1')\cdot
\varepsilon(q_1')]h_{1}^{(\text{v})}(k^{2})(p_1'+q_1')_{\mu}-h^{(\text{v})}_{2}(k^{2})\{[\varepsilon^*(p_1')\cdot
q_1']\varepsilon_{\mu}(q_1')\\
&+[\varepsilon(q_1')\cdot
p_1']\varepsilon^*_{\mu}(p_1')\}-h_{3}^{(\text{v})}(k^{2})\frac{1}{M_1^{2}}[\varepsilon^*(p_1')\cdot
q_1'][\varepsilon(q_1')\cdot p_1'](p_1'+q_1')_{\mu}\bigg\},
\end{split}
\end{equation}
\begin{equation}\label{meofqc2}
\begin{split}
\langle V(p_2')|J_{\nu}|V(q_2')
\rangle=&\frac{1}{2\sqrt{E_2(p_2')E_2(q_2')}}\bigg\{[\varepsilon^*(p_2')\cdot
\varepsilon(q_2')]\bar{h}_{1}^{(\text{v})}(k^{2})(p_2'+q_2')_{\nu}-\bar{h}^{(\text{v})}_{2}(k^{2})\{[\varepsilon^*(p_2')\cdot
q_2']\varepsilon_{\nu}(q_2')\\
&+[\varepsilon(q_2')\cdot
p_2']\varepsilon^*_{\nu}(p_2')\}-\bar{h}_{3}^{(\text{v})}(k^{2})\frac{1}{M_2^{2}}[\varepsilon^*(p_2')\cdot
q_2'][\varepsilon(q_2')\cdot p_2'](p_2'+q_2')_{\nu}\bigg\},
\end{split}
\end{equation}
where $p'_1=(\textbf{p}',ip_{10}')$, $p'_2=(-\textbf{p}',ip_{20}')$,
$q'_1=(\textbf{q}',iq_{10}')$, $q'_2=(-\textbf{q}',iq_{20}')$,
$k=p_1'-q_1'=q_2'-p_2'$ is the momentum of the light meson and
$\textbf{k}=\textbf{p}'-\textbf{q}'$; $h(k^2)$ and $\bar h(k^2)$ are
scalar functions, the four-vector $\varepsilon(p)=(\bm{\varepsilon}+\frac{(\bm{\varepsilon}\cdot\textbf{p})\textbf{p}}{M_H(E_H(p)+M_H)},i\frac{\bm{\varepsilon}\cdot\textbf{p}}{M_H})$
is the polarization vector of heavy vector meson with momentum
$\text{p}$, $E_H(p)=\sqrt{\textbf{p}^2+M_H^2}$,
$(\bm{\varepsilon},0)$ is the polarization vector in the heavy meson
rest frame. In this approach, we calculate the meson-meson interaction when the exchange-meson is
off the mass shell ($k^2\neq -m^2$) and the heavy meson form factors $h(k^2)$ and $\bar
h(k^2)$ are necessarily required. In section \ref{sec:formfac}, we will show how to obtain the
form factors containing the contribution from the vacuum condensates. Taking away the external lines including the
normalizations and polarization vectors $\varepsilon^*_\alpha(p_1'),
\varepsilon_\theta(q_1'), \varepsilon^*_\beta(p_2'),
\varepsilon_{\kappa}(q_2')$ in Eq. (\ref{effinter}), we obtain the interaction kernel from
one light vector meson ($V_1$ and $V_8$) exchange
\begin{equation}\label{kernel}
\begin{split}
\mathcal {V}^I_{\alpha\theta,\beta\kappa}(p',q';P')=&\bigg(\frac{g_1^2}{k^2+m_\omega^2}+\frac{g_8^2}{k^2+m_\phi^2}\bigg)\bar{\mathcal {V}}_{\alpha\theta,\beta\kappa}(p',q';P')\\
=&\bigg(\frac{g_1^2}{k^2+m_\omega^2}+\frac{g_8^2}{k^2+m_\phi^2}\bigg)\{h_1^{(\text{v})}(k^2)\bar{h}_1^{(\text{v})}(k^2)(p_1'+q_1')\cdot(p_2'+q_2')\delta_{\alpha\theta}\delta_{\beta\kappa}\\
&-h_1^{(\text{v})}(k^2)\bar{h}_2^{(\text{v})}(k^2)\delta_{\alpha\theta}[q_{2\beta}'(p_1'+q_1')_\kappa+(p_1'+q_1')_\beta p_{2\kappa}']\\
&-h_2^{(\text{v})}(k^2)\bar{h}_1^{(\text{v})}(k^2)[q_{1\alpha}'(p_2'+q_2')_\theta+(p_2'+q_2')_\alpha
p_{1\theta}']\delta_{\beta\kappa}\\&+h_2^{(\text{v})}(k^2)\bar{h}_2^{(\text{v})}(k^2)[q_{1\alpha}'q_{2\beta}'\delta_{\theta\kappa}
+q_{1\alpha}'\delta_{\theta\beta}p_{2\kappa}'+\delta_{\alpha\kappa}p_{1\theta}'q_{2\beta}'+\delta_{\alpha\beta}p_{1\theta}'p_{2\kappa}']\},
\end{split}
\end{equation}
where $k=(\textbf{k},0)$.

Then using the method above, we can obtain the interaction kernels from one light vector meson ($\rho^{\pm}$ and $K^*$) exchange, as shown in Fig. {\ref{Fig2}}(b), (c), (d), (f), (g), (h),
\begin{equation}\label{kernel1}
\begin{split}
\mathcal {V}^\rho_{\alpha\theta,\beta\kappa}(p',q';P')=\frac{2g_\rho^2}{k^2+m_\rho^2}\bar{\mathcal {V}}_{\alpha\theta,\beta\kappa}(p',q';P'),~~
\mathcal {V}^{K^*}_{\alpha\theta,\beta\kappa}(p',q';P')=\frac{g_{K^*}^2}{k^2+m_{K^*}^2}\bar{\mathcal {V}}_{\alpha\theta,\beta\kappa}(p',q';P'),
\end{split}
\end{equation}
where $\rho$ represents $\rho^+$ and $\rho^-$ mesons, $K^*$ represents $K^{*+}$, $K^{*-}$, $K^{*0}$ and $\bar{K}^{*0}$ mesons, $g_\rho$ and $g_{K^*}$ are the meson-quark coupling constants obtained within QCD sum rules approach, $g_{\rho}^2=2.42$, \cite{cc2} and $g_{K^*}^2=1.46$ \cite{cc3}.

\section{The extended Bethe-Salpeter equation}\label{sec:instanapp}
In this section, we solve the BS equation expressed as Eq. (\ref{BSE1}) with the kernel (\ref{kernel}) in instantaneous approximation and obtain the wave function of the pure molecule state $D^{*+}_s$$D^{*-}_s$. Firstly, we assume that the quantum numbers of pure molecule state $D^{*+}_s$$D^{*-}_s$ are $J^P=0^+$. Substituting its BS wave function given by Eq. (\ref{SBSWF}) and the kernel (\ref{kernel}) into the BS equation (\ref{BSE1}), we find that the integral of one term on the right-hand side of Eq. (\ref{SBSWF}) has contribution to the one of
itself and the other term. But these cross terms have the
factors of $1/M_1^2$ and $1/M_2^2$, which are small
for the heavy meson masses are large. Ref. \cite{mypaper4} has given a simple approach to solve this BS equation as follow. Ignoring the small cross terms, we can obtain two individual equations:
\begin{equation}\label{BSE2}
\mathcal{F}^1_{\lambda\tau}(P'\cdot p',p'^2) =\int
\frac{id^4q'}{(2\pi)^4}\Delta_{F\lambda\alpha}(p_1')\mathcal
{V}_{\alpha\theta,\beta\kappa}(p',q';P')\mathcal{F}^1_{\theta\kappa}(P'\cdot
q',q'^2)\Delta_{F\beta\tau}(p_2'),
\end{equation}
\begin{equation}\label{BSE3}
\mathcal {F}^2_{\lambda\tau}(P'\cdot p',p'^2) =\int
\frac{id^4q'}{(2\pi)^4}\Delta_{F\lambda\alpha}(p_1')\mathcal
{V}_{\alpha\theta,\beta\kappa}(p',q';P')\mathcal{F}^2_{\theta\kappa}(P'\cdot
q',q'^2)\Delta_{F\beta\tau}(p_2'),
\end{equation}
where $\mathcal{F}^1_{\lambda\tau}(P'\cdot p',p'^2)=T^1_{\lambda\tau}\mathcal{F}_1(P'\cdot
p',p'^2)$ and $\mathcal{F}^2_{\lambda\tau}(P'\cdot
p',p'^2)=T^2_{\lambda\tau}\mathcal
{F}_2(P'\cdot p',p'^2)$. Solving these two equations,
respectively, one can obtain two series of eigenfunctions and
eigenvalues. Because the cross terms are small, we take the
BS wave function to be a linear combination of two
eigenstates $\mathcal{F}^{10}_{\lambda\tau}$ and
$\mathcal{F}^{20}_{\lambda\tau}$ corresponding to the lowest energy in
Eqs. (\ref{BSE2}) and (\ref{BSE3}), respectively. Then in the basis provided by
$\mathcal{F}^{10}_{\lambda\tau}(P'\cdot p',p'^2)=T^1_{\lambda\tau}\mathcal{F}_{10}(P'\cdot
p',p'^2)$ and
$\mathcal{F}^{20}_{\lambda\tau}(P'\cdot p',p'^2)=T^2_{\lambda\tau}\mathcal
{F}_{20}(P'\cdot p',p'^2)$, the BS wave function
$\chi^{0^+}_{\lambda\tau}$ is considered as
\begin{equation}\label{BSwfapprox}
\chi^{0^+}_{\lambda\tau}(P',p')=c_1\mathcal{F}^{10}_{\lambda\tau}(P'\cdot
p',p'^2)+c_2\mathcal{F}^{20}_{\lambda\tau}(P'\cdot p',p'^2).
\end{equation}
Substituting Eq. (\ref{BSwfapprox}) into (\ref{BSE1}) and then comparing the tensor structures in the left and right sides, we
obtain an eigenvalue equation
\begin{subequations}\label{eigeneq}
\begin{equation}\label{eigeneqa}
\begin{split}
&c_1\mathcal{F}_{10}(P'\cdot
p',p'^2)=\\
&\frac{1}{p_1'^2+M_1^2-i\epsilon}\frac{1}{p_2'^2+M_2^2-i\epsilon}\bigg\{\int\frac{id^4q'}{(2\pi)^4}\bigg(\frac{g_1^2}{k^2+m_\omega^2}+\frac{g_8^2}{k^2+m_\phi^2}\bigg)\\
&\times\{h_1^{(\text{v})}(k^2)\bar{h}_1^{(\text{v})}(k^2)(p_1'+q_1')\cdot(p_2'+q_2')+2h_1^{(\text{v})}(k^2)\bar{h}_2^{(\text{v})}(k^2)[(q'_1\cdot q'_2)-(q'_1\cdot p'_2)]\\
&+2h_2^{(\text{v})}(k^2)\bar{h}_1^{(\text{v})}(k^2)[(q'_1\cdot
q'_2)-(p'_1\cdot q'_2)]\}c_1\mathcal{F}_{10}(P'\cdot
q',q'^2)\\
&+\int\frac{id^4q'}{(2\pi)^4}\bigg(\frac{g_1^2}{k^2+m_\omega^2}+\frac{g_8^2}{k^2+m_\phi^2}\bigg)\{2h_1^{(\text{v})}(k^2)\bar{h}_2^{(\text{v})}(k^2)[q_1'^2q_2'^2-q_1'^2(p'_2\cdot
q'_2)]\\
&+2h_2^{(\text{v})}(k^2)\bar{h}_1^{(\text{v})}(k^2)[q_1'^2q_2'^2-q_2'^2(p'_1\cdot
q'_1)]\}c_2\mathcal{F}_{20}(P'\cdot
q',q'^2)\bigg\}
\end{split}
\end{equation}
\begin{equation}\label{eigeneqb}
\begin{split}
&c_2\mathcal{F}_{20}(P'\cdot
p',p'^2)=\\
&\frac{1}{p_1'^2+M_1^2-i\epsilon}\frac{1}{p_2'^2+M_2^2-i\epsilon}\bigg\{\int\frac{id^4q'}{(2\pi)^4}\frac{1}{M_1^2p_2'^2}\bigg(\frac{g_1^2}{k^2+m_\omega^2}+\frac{g_8^2}{k^2+m_\phi^2}\bigg)\\
&\times\{h_1^{(\text{v})}(k^2)\bar{h}_2^{(\text{v})}(k^2)[(p'_1\cdot
p'_2)(q'_1\cdot q'_2)-(p'_1\cdot q'_2)(q'_1\cdot
p'_2)]+h_2^{(\text{v})}(k^2)\bar{h}_1^{(\text{v})}(k^2)\\
&\times[p_1'\cdot(p_2'+q_2')q'_2\cdot(q'_1-p_1')-(M_1^2+(p'_1\cdot
q'_1))q_2'\cdot(p_2'+q_2')]\}c_1\mathcal{F}_{10}(P'\cdot
q',q'^2)\\
&+\int\frac{id^4q'}{(2\pi)^4}\frac{1}{M_1^2p_2'^2}\bigg(\frac{g_1^2}{k^2+m_\omega^2}+\frac{g_8^2}{k^2+m_\phi^2}\bigg)\{h_1^{(\text{v})}(k^2)\bar{h}_1^{(\text{v})}(k^2)\\
&\times(p_1'+q_1')\cdot(p_2'+q_2')
q_2'^2[M_1^2+(p_1'\cdot
q_1')-q_1'^2]+h_1^{(\text{v})}(k^2)\bar{h}_2^{(\text{v})}(k^2)[M_1^2(q_1'\cdot q_2')(p_2'\cdot q_2')\\
&-M_1^2q_2'^2(p_2'\cdot
q_1')+q_1'^2q_2'^2(p_1'\cdot p_2')-q_1'^2(p'_1\cdot q'_2)(p'_2\cdot
q'_2)]+h_2^{(\text{v})}(k^2)\bar{h}_1^{(\text{v})}(k^2)\\
&\times q_2'^2[p_1'\cdot(p_2'+q_2')q_1'\cdot(q_1'-p_1')
-(M_1^2+(p_1'\cdot
q_1'))q_1'\cdot(p_2'+q_2')]\}c_2\mathcal{F}_{20}(P'\cdot
q',q'^2)\bigg\}.
\end{split}
\end{equation}
\end{subequations}
From this eigenvalue equation, we can obtain the eigenfunctions and eigenvalues including the contribution from
the cross terms.

In instantaneous approximation, Eqs. (\ref{BSE2}) and (\ref{BSE3}) become the Schr$\ddot{o}$dinger type equations, respectively, (see details in Appendix B)
\begin{equation}\label{BSE5}
\left(\frac{b_1^2(M)}{2\mu_R}-\frac{\textbf{p}'^2}{2\mu_R}\right)\Psi_1^{0^+}(\textbf{p}')=\int
\frac{d^3k}{(2\pi)^3}V_{1}^{0^+}(\textbf{p}',\textbf{k})\Psi_1^{0^+}(\textbf{p}',\textbf{k}),
\end{equation}
\begin{equation}\label{BSE7}
\left(\frac{b_2^2(M)}{2\mu_R}-\frac{\textbf{p}'^2}{2\mu_R}\right)\Psi_2^{0^+}(\textbf{p}')=\int
\frac{d^3k}{(2\pi)^3}V_{2}^{0^+}(\textbf{p}',\textbf{k})\Psi_2^{0^+}(\textbf{p}',\textbf{k}),
\end{equation}
where $\Psi_1^{0^+}(\textbf{p}')=\int dp'_0F_1(P'\cdot p',p'^2)$, $\Psi_2^{0^+}(\textbf{p}')=\int
dp'_0p_2'^2\mathcal{F}_2(P'\cdot p',p'^2)$,
$\mu_R=E_1E_2/(E_1+E_2)=[M^4-(M_1^2-M_2^2)^2]/(4M^3)$,
$b^2(M)=[M^2-(M_1+M_2)^2][M^2-(M_1-M_2)^2]/(4M^2)$,
and the potentials between $D^{*+}_s$ and $D^{*-}_s$ up to the
second order of the $p'/M_H$ expansion are
\begin{equation}\label{potential1}
V_{1}^{0^+}(\textbf{p}',\textbf{k})=
h_1^{(\text{v})}(k^2)\bigg(\frac{g_1^2}{k^2+m_\omega^2}+\frac{g_8^2}{k^2+m_\phi^2}\bigg)\bar{h}_1^{(\text{v})}(k^2)\bigg[-1-\frac{4\textbf{p}'^2+5\textbf{k}^2}{4E_1E_2}\bigg],
\end{equation}
\begin{equation}\label{potential2}
\begin{split}
V_{2}^{0^+}(\textbf{p}',\textbf{k})=&
h_1^{(\text{v})}(k^2)\bigg(\frac{g_1^2}{k^2+m_\omega^2}+\frac{g_8^2}{k^2+m_\phi^2}\bigg)\bar{h}_1^{(\text{v})}(k^2)\bigg[-1-\frac{2\textbf{p}'^2+2\textbf{k}^2}{4M_1^2}-\frac{2\textbf{p}'^2+2\textbf{k}^2}{4E_1E_2}\bigg].
\end{split}
\end{equation}
The eigenvalue equation (\ref{eigeneq}) becomes
\begin{equation}\label{eigeneq2}
\left(\begin{array}{cc}\frac{b_{10}^2(M)}{2\mu_R}-\lambda&H_{12}\\H_{21}&\frac{b_{20}^2(M)}{2\mu_R}-\lambda\end{array}
\right)\left(\begin{array}{c}c'_1\\c'_2\end{array}\right)=0,
\end{equation}
where the matrix elements are
\begin{equation}\label{me1}
\begin{split}
H_{12}=H_{21}=&\int d^3p'\Psi_{10}^{0^+}(\textbf{p}')^*\int\frac{d^3k}{(2\pi)^3}h_1^{(\text{v})}(k^2)\bigg(\frac{g_1^2}{k^2+m_\omega^2}+\frac{g_8^2}{k^2+m_\phi^2}\bigg)\bar{h}_1^{(\text{v})}(k^2)\frac{\textbf{k}^2}{E_1E_2}\Psi_{20}^{0^+}(\textbf{p}',\textbf{k}),
\end{split}
\end{equation}
and $\Psi_{10}^{0^+}$ and $\Psi_{20}^{0^+}$  are the
eigenfunctions corresponding to the lowest energy in Eqs. (\ref{BSE5}) and
(\ref{BSE7}), respectively; $b_{10}^2(M)/(2\mu_R)$ and $b_{20}^2(M)/(2\mu_R)$
are the corresponding eigenvalues. The wave function of the pure molecule state $D^{*+}_s$$D^{*-}_s$ becomes
\begin{equation}\label{BSwfapprox1}
\Psi_{D^{*+}_sD^{*-}_s}^{0^+}(\textbf{p}')=c'_1\Psi_{10}^{0^+}(\textbf{p}')+c'_2\Psi_{20}^{0^+}(\textbf{p}').
\end{equation}
Then we can investigate the alternative $J^P=2^+$
assignment for the pure molecule state $D^{*+}_s$$D^{*-}_s$ using the method as above.

\section{Form factors of heavy vector mesons}\label{sec:formfac}
In our previous works \cite{mypaper4,mypaper,mypaper2,mypaper3} the form factors of heavy vector meson $h(k^2)$ describing its internal structure have been calculated, but we did not consider the nonperturbative effects of QCD. In this work, we firstly improve the heavy vector meson BS wave function so that it can include the nonperturbative contribution. The BS amplitude of heavy vector mesons has the form \cite{BSE:Roberts4}
\begin{equation}
\Gamma_{\mu}^{V}(p;P)=\frac{1}{N^{V}}\bigg(\gamma_{\mu}+P_{\mu}\frac{\gamma\cdot
P}{M_{V}^{2}}\bigg)\varphi_{H}(p^{2}),
\end{equation}
where $P$ is the momentum of the heavy vector meson, $p$ denotes the
relative momentum between quark and antiquark in heavy meson, $N^V$ is
the normalization and $\varphi_{H}(p^{2})$=exp$(-p^{2}/\omega_{H}^{2})$. Because the form factors depend on the momentum of the exchange-meson only, we set $P=(0,0,0,iM_{V})$ in the rest frame to calculate these scalar functions. The authors of Ref. \cite{BSE:Roberts4} considered the SU(3) symmetry and obtained $\omega_D=\omega_{D^{*0}}=\omega_{D^{*+}_s}$=1.81Gev. These parameters are fixed by providing fits to the observables. As in heavy-quark effective theory (HQET) \cite{hqet}, we consider that the heaviest quark carries all the heavy-meson momentum. Let $D^*_l$ denote one of $D^{*0}$, $D^{*+}$ and $D^{*+}_s$, and $l=u,d,s$ represents the $u,d,s$-antiquark in the heavy vector meson $D^{*0}$, $D^{*+}$ and $D^{*+}_s$, respectively. Then the BS wave function of $D^{*}_l$ is obtained
\begin{equation}\label{BSwavefunc}
\chi= S_c(p+P)\Gamma_{\mu}^{V}(p;P)S_l(p),
\end{equation}
where $S_c(p+P)$ is c-quark propagator and $S_l(p)$ is the light quark propagator in constituent quark model.

The operator product expansion (OPE) was introduced to deal with the nonperturbative effects of QCD \cite{ope1}. Its physical meaning is that the short distance behaviour is determined by the Wilson coefficients and the large distance part is included in the matrix elements of the operators $O_n$ \cite{cc2}. Applying the fixed-point gauge technique, the authors of Ref. \cite{cc2} have obtained the massive quark propagators which include the information of condensates
\begin{equation}\label{qcpro}
\begin{split}
S_c(p+P)=&\frac{-i}{\gamma\cdot p+\gamma\cdot P-im_c}+\frac{i}{4}gt^aG^a_{\kappa\lambda}\frac{1}{[(p+P)^2+m_c^2]^2}\\
&\times\{\sigma_{\kappa\lambda}(\gamma\cdot p+\gamma\cdot P+im_c)+(\gamma\cdot p+\gamma\cdot P+im_c)\sigma_{\kappa\lambda}\}\\
&+\frac{i}{12}g^2G^a_{\alpha\beta}G^a_{\alpha\beta}m_c\frac{(p+P)^2+im_c(\gamma\cdot
p+\gamma\cdot P)}{[(p+P)^2+m_c^2]^4},
\end{split}
\end{equation}
and
\begin{equation}\label{qspro}
\begin{split}
S_l(p)=&\frac{-i}{\gamma\cdot p-im_l}+\frac{i}{4}gt^{a'}G^{a'}_{\kappa'\lambda'}\frac{1}{[p^2+m_l^2]^2}\{\sigma_{\kappa'\lambda'}(\gamma\cdot p+im_l)+(\gamma\cdot p+im_l)\sigma_{\kappa'\lambda'}\}\\
&+\frac{i}{12}g^2G^a_{\alpha\beta}G^a_{\alpha\beta}m_l\frac{p^2+im_l\gamma\cdot
p}{(p^2+m_l^2)^4},
\end{split}
\end{equation}
where $m_{c,l}$ are the constituent quark masses,
$\alpha_s=g^2/4\pi$ is the QCD coupling constant, $G^a_{\mu\nu}$ is
the gluon field tensor, $t^a=\lambda^a/2$ and $\lambda^a$ are the
Gell-Mann matrices of the group $SU(3)$, and
$\sigma_{\kappa\lambda}=\frac{1}{2}i[\gamma_\kappa,\gamma_\lambda]$.

Putting the propagators given by Eqs. (\ref{qcpro}) and (\ref{qspro}) into (\ref{BSwavefunc}), one can obtain the BS wave function of heavy vector meson which has the contribution from the vacuum condensates.
The BS wave function expressed as Eq. (\ref{BSwavefunc}) is a $4\times4$ matrix which can be written as a combination of 16 linearly independent matrices $1$, $\gamma_{\mu}$, $\sigma_{\mu\nu}$,
$\gamma_{\mu}\gamma_{5}$, $\gamma_{5}$: \cite{mypaper}
\begin{equation}\label{BSmatrixform}
\chi=\Psi^{S}+\Psi^{V}_{\mu}\gamma_{\mu}+\Psi^{T}_{\mu\nu}\sigma_{\mu\nu}+\Psi^{AV}_{\mu}\gamma_{\mu}\gamma_{5}+\Psi^{Pse}\gamma_{5}.
\end{equation}
The corresponding wave function in instantaneous approximation can
be obtained from Eq. (\ref{BSmatrixform}): the heavy vector meson
wave function is a three-vector, its components are
\cite{mypaper2,mypaper3}
\begin{equation}
\Psi_i^V(\textbf{p})=\int
dp_0\frac{1}{4}tr\{\gamma_i\chi\}~~~i=1,2,3.
\end{equation}
To simplify this integral, we consider that the fourth components of
momenta in the second and third terms in Eqs. (\ref{qcpro}) and
(\ref{qspro}) are equal to zero. Picking up all terms proportional
to $G^2$, we obtain the wave function of $D_l^{*}$ which contains the
nonperturbative contribution from the gluon condensates
\begin{equation}\label{wfV}
\begin{split}
\Psi_{D^{*}_l}(\textbf{p})=&\frac{2\pi
i}{N^V}\bigg\{\frac{1}{4\omega_c\omega_l}\bigg\{exp\bigg[\frac{-2\textbf{p}^2-M_{D^{*}_l}^2-m_c^2+2M_{D^{*}_l}\omega_c+\Gamma^2/4+(M_{D^{*}_l}-\omega_c)i\Gamma}{\omega_{D^*_l}^2}\bigg]\\
&\times\bigg[\frac{-M_{D^{*}_l}+\omega_c-\omega_l-i\Gamma/2}{(M_{D^{*}_l}-\omega_c+\omega_l)^2+\Gamma^2/4}+\frac{M_{D^{*}_l}-\omega_c-\omega_l+i\Gamma/2}{(M_{D^{*}_l}-\omega_c-\omega_l)^2+\Gamma^2/4}\bigg]\bigg[\frac{\textbf{p}^2}{3}+\omega_c
M_{D^{*}_l}-\omega_c^2+\frac{\Gamma^2}{4}\\
&+m_cm_l+i\bigg(\frac{M_{D^{*}_l}\Gamma}{2}-\omega_c\Gamma\bigg)\bigg]-exp\bigg(\frac{-2\textbf{p}^2-m_l^2}{\omega_{D^*_l}^2}\bigg)\bigg(\frac{\textbf{p}^2}{3}-\omega_l^2+M_{D^{*}_l}\omega_l+m_cm_l\bigg)\\
&\times\frac{M_{D^{*}_l}-\omega_c-\omega_l+i\Gamma/2}{(M_{D^{*}_l}-\omega_c-\omega_l)^2+\Gamma^2/4}\bigg\}
+\langle 0|g^2G^a_{\mu\nu}G^a_{\mu\nu}|0 \rangle\bigg\{\frac{1}{72}
exp\bigg(\frac{-\textbf{p}^2}{\omega_{D^*_l}^2}\bigg)\bigg(m_cm_l+\frac{\textbf{p}^2}{3}\bigg)\\
&\times\frac{\sqrt{\textbf{p}^2+m_c^2}+\sqrt{\textbf{p}^2+m_l^2}}{(\textbf{p}^2+m_c^2)^2(\textbf{p}^2+m_l^2)^2}+\frac{i}{96} exp\bigg(\frac{-2\textbf{p}^2-m_l^2}{\omega_{D^*_l}^2}\bigg)\bigg(m_l\textbf{p}^2-m_c\frac{\textbf{p}^2}{3}\bigg)\frac{m_c}{2\omega_l(\textbf{p}^2+m_c^2)^4}\\
&-\frac{i}{96}exp\bigg[\frac{-2\textbf{p}^2-M_{D^{*}_l}^2-m_c^2+2M_{D^{*}_l}\omega_c+\Gamma^2/4+(M_{D^{*}_l}-\omega_c)i\Gamma}{\omega_{D^*_l}^2}\bigg]\bigg(m_c\textbf{p}^2-m_l\frac{\textbf{p}^2}{3}\bigg)\\
&\times\frac{m_l}{2\omega_c(\textbf{p}^2+m_l^2)^4}\bigg\}\bigg\}\sqrt{3}\hat{\textbf{p}},
\end{split}
\end{equation}
where $M_{D^{*}_l}$ is the mass of the heavy meson $D^{*}_l$, $\hat{\textbf{p}}$ is the unit momentum,
$\Gamma$ is the width of resonance, $\omega_{c,l}=\sqrt{\textbf{p}^2+m_{c,l}}$ and $l=u,d,s$. To obtain this wave function, we have used the substitution
\begin{equation}
\langle0|G^a_{\nu\mu}G^b_{\sigma\rho}|0\rangle=\frac{1}{96}\delta_{ab}(g_{\mu\rho}g_{\nu\sigma}-g_{\mu\sigma}g_{\nu\rho})\langle0|G^a_{\mu\nu}G^a_{\mu\nu}|0\rangle.
\end{equation}

\begin{figure*}
\centering
\resizebox{0.5\textwidth}{!}{%
  \includegraphics{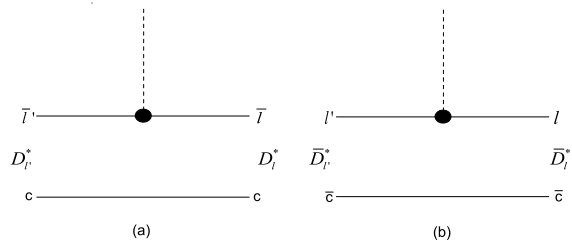}
}
\caption{\label{Fig3} The vertex function $\Gamma$ in the impulse
approximation. Diagram (a) represents the light meson interaction with $l$-antiquark in $D_l^{*}$ and $l'$-antiquark in $D_{l'}^{*}$. Diagram (b) represents the light meson interaction with $l$-quark in $\bar{D}_l^{*}$ and $l'$-quark in $\bar{D}_{l'}^{*}$.}
\end{figure*}

Now, we calculate these form factors derived from one light vector meson (also including $\rho^{\pm}$ and $K^*$) exchange. After exchanging one vector meson, the final particle may not be the initial one, as shown in Fig. \ref{Fig2}. The quark current becomes $J_\mu=i\bar{l'}\gamma_\mu l$, where $l'=u,d,s$ and $l$ represent the light quarks in final and initial particles, respectively. The matrix element of the quark current $J_\mu$ between the heavy meson states (H) has the form
\cite{FF:Faustov1,FF:Faustov2}
\begin{equation} \label{meH}
\langle H(Q)|J_\mu(0)|H(P) \rangle=\int
\frac{d^{3}qd^{3}p}{(2\pi)^{6}}\bar{\Psi}_{Q}^{H}(\textbf{q})\Gamma_\mu(\textbf{p},\textbf{q})\Psi_{P}^{H}(\textbf{p}),
\end{equation}
where $\Gamma_\mu(\textbf{p},\textbf{q})$ is the two-particle vertex
function and $\Psi_{Q}^{H}$ is the wave function of heavy vector
meson projected onto the positive energy states of quarks and
boosted to the moving reference frame with momentum Q. Fig. \ref{Fig3} shows the vertex
function $\Gamma_\mu(\textbf{p},\textbf{q})$ in the impulse
approximation. The corresponding vertex
function of the quark-meson interaction is given by
\begin{equation} \label{vertex}
\Gamma_\mu^{(1)}(\textbf{p},\textbf{q})=\left\{ \begin{array}{cc} \bar{v}_{\bar{l'}}(q_{1})i\gamma_\mu v_{\bar{l}}(p_{1})(2\pi)^{3}\delta(\textbf{q}_{2}-\textbf{p}_{2})   \\
\bar{u}_{l'}(q_{1})i\gamma_\mu u_{l}(p_{1})(2\pi)^{3}\delta(\textbf{q}_{2}-\textbf{p}_{2})
\end{array} \right.,
\end{equation}
where $u_l(p)$ and $v_{\bar{l}}(p)$ are the spinors of the quark $l$
and antiquark $\bar{l}$, respectively,
\begin{equation}
u_\lambda(p)=\sqrt{\frac{\epsilon_l(p)+m_l}{2\epsilon_l(p)}} \left( \begin{array}{c} 1 \\
\frac{ {\bf \sigma}\cdot {\bf p}}{\epsilon_l(p)+m_l}
\end{array} \right) \chi_{\lambda},~v_\lambda(p)=\sqrt{\frac{\epsilon_l(p)+m_l}{2\epsilon_l(p)}}
\left( \begin{array}{c} \frac{ {\bf \sigma}\cdot {\bf p}}{ \epsilon_l(p)+m_l} \\
1
\end{array} \right) \chi_{\lambda},
\end{equation}
with $\epsilon_{l,c}(p)=\sqrt{\textbf{p}^{2}+m_{l,c}^{2}}$   and
\cite{FF:Faustov1,FF:Faustov2}
\begin{equation*}
\begin{split}
p_{1,2}&=\epsilon_{l,c}(p)\frac{P}{M_{H}}\pm\sum^{3}_{i=1}n^{(i)}(P)p_{i},~~M_{H}=\epsilon_{l}(p)+\epsilon_{c}(p),\\
q_{1,2}&=\epsilon_{l,c}(q)\frac{Q}{M_{H}}\pm\sum^{3}_{i=1}n^{(i)}(Q)q_{i},~~M_{H}=\epsilon_{l}(q)+\epsilon_{c}(q),
\end{split}
\end{equation*}
and $n^{(i)}$ are three four-vectors defined by
\begin{equation*}
n^{(i)}(P)=\left\{\delta_{ij}+\frac{P_iP_j}{M_{H}[E_{H}(P)+M_{H}]},~i\frac{P_i}{M_{H}} \right\},~~
E_{H}(P)=\sqrt{\textbf{P}^{2}+M_{H}^{2}}.
\end{equation*}
The first term in Eq. (\ref{vertex}) represents the light vector
meson interaction with the $l$-antiquark in $D_l^{*}$ and $l'$-antiquark in $D_{l'}^{*}$, while the second
term is its interaction with the $l$-quark in $\bar{D}_l^{*}$ and $l'$-quark in $\bar{D}_{l'}^{*}$, where $\bar{D}_l^{*}$ denotes the anti-particle of $D_l^{*}$.

\begin{figure*}
\centering
\resizebox{0.5\textwidth}{!}{%
  \includegraphics{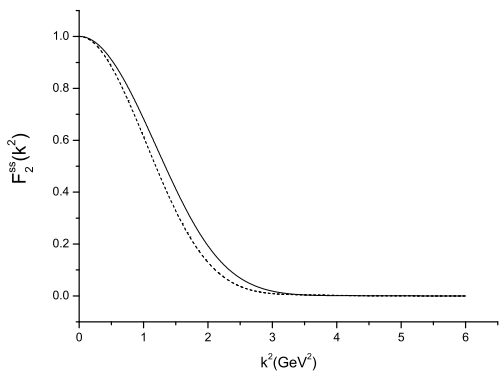}
}
\caption{\label{Fig4} The form factor for the vertex of heavy vector
meson $D^{*}_s$ coupling to light vector ($V_1$ and $V_8$) meson. The solid line represents the form factor including the contribution from gluon condensates; and the dashed line represents the form factor without this contribution.}
\end{figure*}

Substituting the vertex function $\Gamma_\mu^{(1)}$ given by Eq.
(\ref{vertex}) into the matrix element (\ref{meH}) and comparing the
resulting expressions with the form factor decompositions
(\ref{meofqc1}) and (\ref{meofqc2}), we obtain
\begin{equation}\label{ffv}
\begin{split}
h_{1}^{(\text{v})}(k^{2})=&h_{2}^{(\text{v})}(k^{2})=\bar{h}_{1}^{(\text{v})}(k^{2})=\bar{h}_{2}^{(\text{v})}(k^{2})=F^{l'l}_2({\textbf{k}^{2}}),~~h_{3}^{(\text{v})}(k^{2})=\bar{h}_{3}^{(\text{v})}(k^{2})=0,\\
F^{l'l}_2(\textbf{k}^{2})=&\frac{2\sqrt{E_{D^*_{l'}}M_{D^*_{l}}}}{E_{D^*_{l'}}+M_{D^*_{l}}}\int
\frac{d^{3}p}{(2\pi)^{3}}\bar{\Psi}_{D^{*}_{l'}}(\textbf{p}+\frac{2\epsilon_c(p)}{E_{D^*_{l'}}+M_{D^*_{l'}}}\textbf{k})\sqrt{\frac{\epsilon_l(p)+m_l}{\epsilon_{l'}(p+k)+m_{l'}}}\\
&\times\left\{ \frac{\epsilon_{l'}(p+k)+\epsilon_l(p)+m_{l'}-m_l}{2
\sqrt{\epsilon_{l'}(p+k)
\epsilon_l(p)}}+\frac{\textbf{pk}}{2\sqrt{\epsilon_{l'}(p+k)\epsilon_l(p)}(\epsilon_l(p)+m_l)}
\right\}\Psi_{D^{*}_l}(\textbf{p}),
\end{split}
\end{equation}
where $\Psi_{D^{*}_l}(\textbf{p})$ is the heavy vector meson wave function expressed as Eq. (\ref{wfV}). In Fig. \ref{Fig4} we give the form factor $F^{ss}_{2}(\textbf{k}^{2})$ of heavy vector meson $D^*_s$ corresponding to one light vector meson ($V_1$ and $V_8$) exchange and compare it with the form factor without the contribution from gluon condensates.

Finally, the potentials given by Eqs. (\ref{potential1}) and (\ref{potential2}) become
\begin{equation}\label{potential3}
\begin{split}
V_{1}^{0^+}(\textbf{p}',\textbf{k})=&-F^{ss}_2(\textbf{k}^2)\bigg(\frac{g_1^2}{k^2+m_\omega^2}+\frac{g_8^2}{k^2+m_\phi^2}\bigg)
F^{ss}_2(\textbf{k}^2)\bigg[1+\frac{4\textbf{p}'^2+5\textbf{k}^2}{4E_1E_2}\bigg],\\
V_{2}^{0^+}(\textbf{p}',\textbf{k})=&-F^{ss}_2(\textbf{k}^2)\bigg(\frac{g_1^2}{k^2+m_\omega^2}+\frac{g_8^2}{k^2+m_\phi^2}\bigg)F^{ss}_2(\textbf{k}^2)\bigg[1+\frac{2\textbf{p}'^2+2\textbf{k}^2}{4M_1^2}+\frac{2\textbf{p}'^2+2\textbf{k}^2}{4E_1E_2}\bigg].
\end{split}
\end{equation}

\section{Numerical result}\label{sec:nr}
The constituent quark masses $m_{c}=1.55$GeV, $m_{u}=m_{d}=0.33$GeV, $m_{s}=0.5$GeV, the
meson masses $m_{\omega}=0.782$GeV, $m_{\rho^0}=m_{\rho^{\pm}}=0.775$GeV,
$m_{K^{*0}}=m_{\bar{K}^{*0}}=0.896$GeV, $m_{K^{*+}}=m_{K^{*-}}=0.892$GeV, $m_{\phi}=1.019$GeV, $m_{D^{*+}_s}=m_{D^{*-}_s}=2.112$GeV, the width
of the heavy vector meson $\Gamma_{D^*_s}=0.0015$GeV \cite{PDG2014}
and the gluon condensate
$\langle0|\frac{\alpha_s}{\pi}G^a_{\mu\nu}G^a_{\mu\nu}|0\rangle=(0.36GeV)^4$
\cite{cc2}. With these potentials expressed as Eq.
(\ref{potential3}), we can numerically solve the equations
(\ref{BSE5}) and (\ref{BSE7}), respectively. Subsequently, the
eigenvalue equation (\ref{eigeneq2}) can be solved. In this
approach, we can investigate the molecular state $D^{*+}_sD^{*-}_s$
with $J^P=2^+$. The obtained ground-state masses of the pure molecule state $D_s^{*+}D_s^{*-}$ with $J^P=0^+, 2^+$ are
presented in Table \ref{table1}. In Ref. \cite{mypaper4}, we have considered one light meson ($\sigma,\rho^0,V_1,V_8$) exchange and obtained the masses and wave functions of pure molecule states $D^{*0}\bar{D}^{*0}$ and $D^{*+}D^{*-}$ collected in Table \ref{table1}.

\begin{table}[ht]
\caption{The ground-state masses of pure molecule states (in
GeV).} \label{table1} \vspace*{-6pt}
\begin{center}
\begin{tabular}{ccc}\hline
Composite & ~~~ State $J^{P}$ &~~~~ This work
\\ \hline
   $D^{*0}\bar{D}^{*0}$  & $0^{+}$ & 3.948
\\   & $2^{+}$ &  4.147
\\  $D^{*+}D^{*-}$  & $0^{+}$ & 3.953
\\   & $2^{+}$ &  4.153\\
$D_s^{*+}D_s^{*-}$  & $0^{+}$ & 4.184
\\   & $2^{+}$ &  4.341
\\
\hline
\end{tabular}
\end{center}
\end{table}

The meson-quark coupling constants $g_\omega$, $g_\rho$, $g_\phi$, $g_{K^*}$ and the parameters $\omega_H$ in the BS amplitude of heavy vector mesons are fixed by providing fits to observables, so there is not an adjustable  parameter in our approach. Simultaneously varying the constituent quark masses $m_s$, $m_c$, and the full width of
the heavy vector meson $\Gamma_{D_s^*}$ within 5\%, we find that the ratio of the numerical result difference dependent on these
parameters and the binding energy is at most 5\%. Thus in our approach
the calculated masses of pure molecule state depend on these parameters,
but not sensitively.

For the pure molecule states $D^{*0}\bar{D}^{*0}$, $D^{*+}D^{*-}$ and $D_s^{*+}D_s^{*-}$, we find that the resulting ground states with $0^+$ lie slightly below and the $2^+$ states above the threshold and then consider that the \emph{Y}(4140) is a mixed state of these three ground states with $J^P=0^+$. From Eq. (\ref{mixBSwf}), we can obtain the wave function of the mixed state in instantaneous approximation
\begin{equation}
\Psi=a'_1\Psi_{D^{*0}\bar{D}^{*0}}+a'_2\Psi_{D^{*+}D^{*-}}+a'_3\Psi_{D_s^{*+}D_s^{*-}},
\end{equation}
where $\Psi_{D_s^{*+}D_s^{*-}}$ is the ground-state wave function of pure molecule state $D_s^{*+}D_s^{*-}$ given in Eq. (\ref{BSwfapprox1}), $\Psi_{D^{*0}\bar{D}^{*0}}$ and $\Psi_{D^{*+}D^{*-}}$ are the ground-state wave functions of pure molecule states
$D^{*0}\bar{D}^{*0}$ and $D^{*+}D^{*-}$, respectively. Then the coupled equation can be obtained
\begin{equation}\label{coupledeq}
\left(\begin{array}{ccc}\frac{b^2(M_{D^{*0}\bar{D}^{*0}})}{2\mu_R}&A_{12}&A_{13}\\A_{21}&\frac{b^2(M_{D^{*+}D^{*-}})}{2\mu_R}&A_{23}\\A_{31}&A_{32}&\frac{b^2(M_{D_s^{*+}D_s^{*-}})}{2\mu_R}\end{array}
\right)\left(\begin{array}{c}a'_1\\a'_2\\a'_3\end{array}\right)=E\left(\begin{array}{c}a'_1\\a'_2\\a'_3\end{array}\right),
\end{equation}
where $M_{D^{*0}\bar{D}^{*0}}$, $M_{D^{*+}D^{*-}}$ and $M_{D_s^{*+}D_s^{*-}}$ are the ground-state masses for the pure molecule states with $J^P=0^+$ given in Table \ref{table1}, $A_{12}$, $A_{13}$, $A_{21}$, $A_{23}$, $A_{31}$ and $A_{32}$
are the matrix elements between two pure molecule states corresponding to the graphs (b), (c), (d), (f), (g), (h) in Fig. \ref{Fig2}, respectively. From the interaction kernel derived from one-$\rho^{\pm}$ exchange given in Eq. (\ref{kernel1}), we obtain these matrix elements at leading order in $p'/M_H$
\begin{equation}\label{mixme1}
A_{ij}=\int d^3p'\Psi^*_{D_{l'}^{*}\bar{D}_{l'}^{*}}(\textbf{p}')\int\frac{d^3k}{(2\pi)^3}F^{l'l}_2(\textbf{k}^{2})\frac{g_{\rho}^2}{k^2+m_{\rho}^2}F^{l'l}_2(\textbf{k}^{2})\Psi_{D_l^{*}\bar{D}_l^{*}}(\textbf{p}',\textbf{k}),
\end{equation}
where we have for $i=1,j=2$, $\rho=\rho^+$, $l'=d$ and $l=u$; for $i=2,j=1$, $\rho=\rho^-$, $l'=u$ and $l=d$. And for one-$K^*$ exchange, the matrix elements become
\begin{equation}\label{mixme2}
A_{ij}=\int d^3p'\Psi^*_{D_{l'}^{*}\bar{D}_{l'}^{*}}(\textbf{p}')\int\frac{d^3k}{(2\pi)^3}F^{l'l}_2(\textbf{k}^{2})\frac{g_{K^*}^2}{k^2+m_{K^*}^2}F^{l'l}_2(\textbf{k}^{2})\Psi_{D_l^{*}\bar{D}_l^{*}}(\textbf{p}',\textbf{k}),
\end{equation}
where we have for $i=1,j=3$, $K^*=K^{*+}$, $l'=s$ and $l=u$; for $i=2,j=3$, $K^*=K^{*0}$, $l'=s$ and $l=d$; for $i=3,j=1$, $K^*=K^{*-}$, $l'=u$ and $l=s$; for $i=3,j=2$, $K^*=\bar{K}^{*0}$, $l'=d$ and $l=s$. In Eqs. (\ref{mixme1}) and (\ref{mixme2}), the high order terms of $p'/M_H$ are not considered for the heavy meson masses are large.

\begin{table}[ht]
\caption{Masses and channel probabilities of mixed states.} \label{table2} \vspace*{-6pt}
\begin{center}
\begin{tabular}{cccc}\hline
Mass (GeV) & ~~~$D^{*0}\bar D^{*0}$  &~~~~$D^{*+}D^{*-}$ & $D_s^{*+}D_s^{*-}$
\\ \hline 3.945  & 45\% & 47\% &8\%
\\ 3.960& 51\%  & 48.9\% &  0.1\% 
\\ 4.146&10\% &10\% & 80\% \\
\hline
\end{tabular}
\end{center}
\end{table}

Solving the equation (\ref{coupledeq}), we obtain three eigenstates and the masses and channel probabilities for these eigenstates are presented in Table \ref{table2}. The resulting highest energy of these eigenstates is in good agreement with the experimental mass of the \emph{Y}(4140) state, while the mass of \emph{Y}(4140) is measured to be $4.143$GeV \cite{T.Aaltonen}. Then we consider that this eigenstate represents the \emph{Y}(4140) state and the component $D_s^{*+}D_s^{*-}$ clearly dominates with a 80\% probability in this mixed state. For two other eigenstates, the component $D_s^{*+}D_s^{*-}$ is less than 10\% probability and we consider that these two mixed states approximately belong to an isospin doublet corresponding the exotic state \emph{Y}(3940), while the mass of \emph{Y}(3940) is 3.943GeV in experiment \cite{SK.CHOI}. The interpretation for the \emph{Y}(3940) state is different from the one in Ref. \cite{mypaper4}, this is because the coupled channel has not been considered in Ref. \cite{mypaper4}.

\section{Conclusion}\label{sec:concl}
For the SU(3) symmetry, the exotic state \emph{Y}(4140) is considered as a mixed state of three pure molecule states $D^{*0}$$\bar{D}^{*}$, $D^{*+}$$D^{*-}$ and $D^{*+}_s$$D^{*-}_s$. Applying the
general formalism of the BS wave functions for the bound states consisting of two vector fields, we investigate these pure molecule states. In this work, we introduce the gluon condensates into the BS wave function of the heavy meson and obtain the heavy meson form factors and the interaction between two heavy mesons including the contribution from the nonperturbative effects of QCD, which is different from our previous works. Then using the coupled-channel approach, we obtain the masses and wave functions for the mixed states of these pure molecule states with $J^P=0^+$, which are in good agreement with experimental masses of the \emph{Y}(3940) and \emph{Y}(4140) states. Thus we can conclude that a mixing of pure molecule states should be a more credible candidate to explain the \emph{Y}(4140) state.

\appendix

\section*{Appendix A: The tensor structures in the general form of the BS wave functions}\label{app}
The tensor structures in Eqs. (\ref{jp0}), (\ref{jp}), (\ref{jm0}), (\ref{jm}) are given below \cite{mypaper4}

\begin{equation*}
T_{\lambda\tau}^1=(\eta_1\eta_2P'^2-\eta_1P'\cdot p'+\eta_2P'\cdot p'-p'^2)g_{\lambda\tau}-(\eta_1\eta_2P'_{\lambda}P'_{\tau}+\eta_2P'_{\lambda}p'_{\tau}-\eta_1p'_{\lambda}P'_{\tau}-p'_{\lambda}p'_{\tau}),
\end{equation*}
\begin{equation*}
\begin{split}
T_{\lambda\tau}^2=&(\eta_1^2P'^2+2\eta_1P'\cdot p'+p'^2)(\eta_2^2P'^2-2\eta_2P'\cdot p'+p'^2)g_{\lambda\tau}\\
&+(\eta_1\eta_2P'^2-\eta_1P'\cdot p'+\eta_2P'\cdot p'-p'^2)(\eta_1\eta_2P'_{\lambda}P'_{\tau}-\eta_1P'_{\lambda}p'_{\tau}+\eta_2p'_{\lambda}P'_{\tau}-p'_{\lambda}p'_{\tau})\\
&-(\eta_2^2P'^2-2\eta_2P'\cdot p'+p'^2)(\eta_1 ^2P'_{\lambda}P'_{\tau}+\eta_1P'_{\lambda}p'_{\tau}+\eta_1p'_{\lambda}P'_{\tau}+p'_{\lambda}p'_{\tau})\\
&-(\eta_1^2P'^2+2\eta_1P'\cdot p'+p'^2)(\eta_2 ^2P'_{\lambda}P'_{\tau}-\eta_2P'_{\lambda}p'_{\tau}-\eta_2p'_{\lambda}P'_{\tau}+p'_{\lambda}p'_{\tau}),
\end{split}
\end{equation*}
\begin{equation*}
\begin{split}
T_{\lambda\tau}^3=&\frac{1}{j!}p'_{\{\mu_2}\cdots
p'_{\mu_j}g_{\mu_1\}\lambda}(\eta_1^2P'^2+2\eta_1P'\cdot p'+p'^2)[(\eta_2^2P'^2-2\eta_2P'\cdot p'+p'^2)(\eta_1P'+p')_{\tau}\\
&-(\eta_1\eta_2P'^2-\eta_1P'\cdot p'+\eta_2P'\cdot p'-p'^2)(\eta_2P'-p')_{\tau}]\\
&-p'_{\mu_1}\cdots
p'_{\mu_j}[(\eta_2^2P'^2-2\eta_2P'\cdot p'+p'^2)(\eta_1 ^2P'_{\lambda}P'_{\tau}+\eta_1P'_{\lambda}p'_{\tau}+\eta_1p'_{\lambda}P'_{\tau}+p'_{\lambda}p'_{\tau})\\
&-(\eta_1\eta_2P'^2-\eta_1P'\cdot p'+\eta_2P'\cdot p'-p'^2)(\eta_1\eta_2P'_{\lambda}P'_{\tau}-\eta_1P'_{\lambda}p'_{\tau}+\eta_2p'_{\lambda}P'_{\tau}-p'_{\lambda}p'_{\tau})],
\end{split}
\end{equation*}
\begin{equation*}
\begin{split}
T_{\lambda\tau}^4=&\frac{1}{j!}p'_{\{\mu_2}\cdots
p'_{\mu_j}g_{\mu_1\}\tau}(\eta_2^2P'^2-2\eta_2P'\cdot p'+p'^2)[(\eta_1\eta_2P'^2-\eta_1P'\cdot p'\\
&+\eta_2P'\cdot p'-p'^2)(\eta_1P'+p')_{\lambda}-(\eta_1^2P'^2+2\eta_1P'\cdot p'+p'^2)(\eta_2P'-p')_{\lambda}]\\
&-p'_{\mu_1}\cdots
p'_{\mu_j}[(\eta_1^2P'^2+2\eta_1P'\cdot p'+p'^2)(\eta_2 ^2P'_{\lambda}P'_{\tau}-\eta_2P'_{\lambda}p'_{\tau}-\eta_2p'_{\lambda}P'_{\tau}+p'_{\lambda}p'_{\tau})\\
&-(\eta_1\eta_2P'^2-\eta_1P'\cdot p'+\eta_2P'\cdot p'-p'^2)(\eta_1\eta_2P'_{\lambda}P'_{\tau}-\eta_1P'_{\lambda}p'_{\tau}+\eta_2p'_{\lambda}P'_{\tau}-p'_{\lambda}p'_{\tau})],
\end{split}
\end{equation*}
\begin{equation*}
\begin{split}
T_{\lambda\tau}^5=&(\eta_2P'\cdot p'-\eta_1P'\cdot p'-2p'^2)p'_{\{\mu_2}\cdots
p'_{\mu_j}\epsilon_{\mu_1\}\lambda\tau\xi}p'_\xi\\
&+(2\eta_1\eta_2P'\cdot
p'+\eta_2p'^2-\eta_1p'^2)p'_{\{\mu_2}\cdots
p'_{\mu_j}\epsilon_{\mu_1\}\lambda\tau\xi}P'_\xi\\
&+p'_{\{\mu_2}\cdots
p'_{\mu_j}\epsilon_{\mu_1\}\lambda\xi\zeta}p'_\xi P'_\zeta
p'_\tau+p'_{\{\mu_2}\cdots
p'_{\mu_j}\epsilon_{\mu_1\}\tau\xi\zeta}p'_\xi P'_\zeta
p'_\lambda,
\end{split}
\end{equation*}
\begin{equation*}\
\begin{split}
T_{\lambda\tau}^6=&(P'\cdot p')p'_{\{\mu_2}\cdots
p'_{\mu_j}\epsilon_{\mu_1\}\lambda\tau\xi}p'_\xi-p'^2p'_{\{\mu_2}\cdots
p'_{\mu_j}\epsilon_{\mu_1\}\lambda\tau\xi}P'_\xi\\
&+p'_{\{\mu_2}\cdots
p'_{\mu_j}\epsilon_{\mu_1\}\lambda\xi\zeta}p'_\xi P'_\zeta
p'_\tau-p'_{\{\mu_2}\cdots
p'_{\mu_j}\epsilon_{\mu_1\}\tau\xi\zeta}p'_\xi P'_\zeta
p'_\lambda,
\end{split}
\end{equation*}
\begin{equation*}
\begin{split}
T_{\lambda\tau}^7=&(\eta_2P'^2-\eta_1P'^2-2P'\cdot p')p'_{\{\mu_2}\cdots
p'_{\mu_j}\epsilon_{\mu_1\}\lambda\tau\xi}p'_\xi\\
&+(2\eta_1\eta_2P'^2+\eta_2P'\cdot p'-\eta_1P'\cdot p')p'_{\{\mu_2}\cdots
p'_{\mu_j}\epsilon_{\mu_1\}\lambda\tau\xi}P'_\xi\\
&+p'_{\{\mu_2}\cdots
p'_{\mu_j}\epsilon_{\mu_1\}\lambda\xi\zeta}p'_\xi P'_\zeta
P'_\tau+p'_{\{\mu_2}\cdots
p'_{\mu_j}\epsilon_{\mu_1\}\tau\xi\zeta}p'_\xi P'_\zeta
P'_\lambda,
\end{split}
\end{equation*}
\begin{equation*}
\begin{split}
T_{\lambda\tau}^8=&P'^2p'_{\{\mu_2}\cdots
p'_{\mu_j}\epsilon_{\mu_1\}\lambda\tau\xi}p'_\xi-(P'\cdot p')p'_{\{\mu_2}\cdots
p'_{\mu_j}\epsilon_{\mu_1\}\lambda\tau\xi}P'_\xi\\
&+p'_{\{\mu_2}\cdots
p'_{\mu_j}\epsilon_{\mu_1\}\lambda\xi\zeta}p'_\xi P'_\zeta
P'_\tau-p'_{\{\mu_2}\cdots
p'_{\mu_j}\epsilon_{\mu_1\}\tau\xi\zeta}p'_\xi P'_\zeta
P'_\lambda.
\end{split}
\end{equation*}

\section*{Appendix B: The instantaneous approximation}
Next, we give the details of the instantaneous approximation. This approach has been used in Ref. \cite{mypaper4}. Comparing the terms $(\eta_1\eta_2P'_{\lambda}P'_{\tau}+\eta_2P'_{\lambda}p'_{\tau}-\eta_1p'_{\lambda}P'_{\tau}-p'_{\lambda}p'_{\tau})$ in both
sides of Eq. (\ref{BSE2}), we obtain
\renewcommand\theequation{B1}
\begin{equation}\label{B1}
\mathcal{F}_1(P'\cdot p',p'^2)
=\frac{1}{p_1'^2+M_1^2-i\epsilon}\frac{1}{p_2'^2+M_2^2-i\epsilon}
\int\frac{id^4q'}{(2\pi)^4}V_{1}(p',q';P')\mathcal{F}_1(P'\cdot
q',q'^2),
\end{equation}
where $V_{1}(p',q';P')$ contains all coefficients of the term
$(\eta_1\eta_2P'_{\lambda}P'_{\tau}+\eta_2P'_{\lambda}p'_{\tau}-\eta_1p'_{\lambda}P'_{\tau}-p'_{\lambda}p'_{\tau})$ in the right side of Eq. (\ref{BSE2}). In
this paper, we set $k=(\textbf{k},0)$ and then
$p'_{10}=q'_{10}=E_1(p_1')=E_1(q_1')$,
$p'_{20}=q'_{20}=E_2(p_2')=E_2(q_2')$. To simplify the potential, we
replace the heavy meson energies $E_1(p_1')=E_1(q_1')\rightarrow
E_1=(M^2-M_2^2+M_1^2)/(2M)$, $E_2(p_2')=E_2(q_2')\rightarrow
E_2=(M^2-M_1^2+M_2^2)/(2M)$. The potential depends on the
three-vector momentum $V(p',q';P')\Rightarrow
V(\textbf{p}',\textbf{q}',M)$. Integrating both sides of Eq.
(\ref{B1}) over $p_0'$ and multiplying by
$(M+\omega_1+\omega_2)(M^2-(\omega_1-\omega_2)^2)$, we obtain
\begin{equation*}
\left(\frac{b_1^2(M)}{2\mu_R}-\frac{\textbf{p}'^2}{2\mu_R}\right)\Psi_1^{0^+}(\textbf{p}')=\int
\frac{d^3k}{(2\pi)^3}V_{1}^{0^+}(\textbf{p}',\textbf{k})\Psi_1^{0^+}(\textbf{p}',\textbf{k})
\end{equation*}
and the potential between $D^{*+}_s$ and $D^{*-}_s$ up to the
second order of the $p'/M_H$ expansion
\begin{equation*}
V_{1}^{0^+}(\textbf{p}',\textbf{k})=
h_1^{(\text{v})}(k^2)\bigg(\frac{g_1^2}{k^2+m_\omega^2}+\frac{g_8^2}{k^2+m_\phi^2}\bigg)\bar{h}_1^{(\text{v})}(k^2)\bigg[-1-\frac{4\textbf{p}'^2+5\textbf{k}^2}{4E_1E_2}\bigg],
\end{equation*}
where $\Psi_1^{0^+}(\textbf{p}')=\int dp'_0F_1(P'\cdot p',p'^2)$,
$\mu_R=E_1E_2/(E_1+E_2)=[M^4-(M_1^2-M_2^2)^2]/(4M^3)$,
$b^2(M)=[M^2-(M_1+M_2)^2][M^2-(M_1-M_2)^2]/(4M^2)$,
$\omega_1=\sqrt{\textbf{p}'^2+M_1^2}$ and
$\omega_2=\sqrt{\textbf{p}'^2+M_2^2}$. And comparing the terms
$(\eta_1 ^2P'_{\lambda}P'_{\tau}+\eta_1P'_{\lambda}p'_{\tau}+\eta_1p'_{\lambda}P'_{\tau}+p'_{\lambda}p'_{\tau})$ in both sides of Eq. (\ref{BSE3}), we
obtain
\begin{equation*}
p_2'^2\mathcal{F}_2(P'\cdot p',p'^2)
=\frac{1}{p_1'^2+M_1^2-i\epsilon}\frac{1}{p_2'^2+M_2^2-i\epsilon}
\int\frac{id^4q'}{(2\pi)^4}V_{2}(p',q';P')q_2'^2\mathcal{F}_2(P'\cdot
q',q'^2).
\end{equation*}
Setting $\Psi_2^{0^+}(\textbf{p}')=\int
dp'_0p_2'^2\mathcal{F}_2(P'\cdot p',p'^2)$, we obtain the equation of
Schr$\ddot{o}$dinger type
\begin{equation*}
\left(\frac{b_2^2(M)}{2\mu_R}-\frac{\textbf{p}'^2}{2\mu_R}\right)\Psi_2^{0^+}(\textbf{p}')=\int
\frac{d^3k}{(2\pi)^3}V_{2}^{0^+}(\textbf{p}',\textbf{k})\Psi_2^{0^+}(\textbf{p}',\textbf{k})
\end{equation*}
and the potential between $D^{*+}_s$ and $D^{*-}_s$ up to the
second order of the $p'/M_H$ expansion
\begin{equation*}
V_{2}^{0^+}(\textbf{p}',\textbf{k})=
h_1^{(\text{v})}(k^2)\bigg(\frac{g_1^2}{k^2+m_\omega^2}+\frac{g_8^2}{k^2+m_\phi^2}\bigg)\bar{h}_1^{(\text{v})}(k^2)\nonumber\bigg[-1-\frac{2\textbf{p}'^2+2\textbf{k}^2}{4M_1^2}-\frac{2\textbf{p}'^2+2\textbf{k}^2}{4E_1E_2}\bigg].
\end{equation*}
Then the eigenfunctions in Eqs.
(\ref{BSE2}) and (\ref{BSE3}) can be calculated in instantaneous approximation and we obtain the
eigenfunctions $\Psi_{10}^{0^+}$ and $\Psi_{20}^{0^+}$ corresponding to the lowest energy in Eqs. (\ref{BSE5}) and
(\ref{BSE7}), respectively; $b_{10}^2(M)/(2\mu_R)$ and $b_{20}^2(M)/(2\mu_R)$
are the corresponding eigenvalues. After integrating both sides of Eqs.
(\ref{eigeneqa}) and (\ref{eigeneqb}) over $p_0'$, we multiply the resulting expressions of (\ref{eigeneqa}) and (\ref{eigeneqb}) from the left by the eigenfunctions $\Psi_{10}^{0^+}$ and $\Psi_{20}^{0^+}$, respectively, and then integrate both sides over the relative momentum $\textbf{p}'$. The eigenvalue equation (\ref{eigeneq}) becomes
\begin{equation*}
\left(\begin{array}{cc}\frac{b_{10}^2(M)}{2\mu_R}-\lambda&H_{12}\\H_{21}&\frac{b_{20}^2(M)}{2\mu_R}-\lambda\end{array}
\right)\left(\begin{array}{c}c'_1\\c'_2\end{array}\right)=0,
\end{equation*}
where the matrix elements are
\begin{equation*}
H_{12}=H_{21}=\int d^3p'\Psi_{10}^{0^+}(\textbf{p}')^*\int\frac{d^3k}{(2\pi)^3}h_1^{(\text{v})}(k^2)\bigg(\frac{g_1^2}{k^2+m_\omega^2}+\frac{g_8^2}{k^2+m_\phi^2}\bigg)\bar{h}_1^{(\text{v})}(k^2)\frac{\textbf{k}^2}{E_1E_2}\Psi_{20}^{0^+}(\textbf{p}',\textbf{k}).
\end{equation*}

\end{document}